\documentclass[sigconf]{acmart}

\usepackage{booktabs} % For formal tables

\setcopyright{acmcopyright}
\acmDOI{xx.xxx/xxx_x}

\acmISBN{978-1-4503-9517-5/23/03}

\acmConference[SAC'23]{ACM SAC Conference}{March 27 –April 2, 2023}{Tallinn, Estonia}
\acmYear{2023}
\copyrightyear{2023}

\acmArticle{4}
\acmPrice{15.00}

\begin{document}
\title{AskYourDB: An end-to-end system for querying and visualizing relational databases using natural language}
% \titlenote{Produces the permission block, and
% \subtitle{Extended Abstract}
% \subtitlenote{The full version of the author's guide is available as
%   \texttt{acmart.pdf} document}
  
\renewcommand{\shorttitle}{AskYourDB}

\author{Manu Joseph}
\authornote{These authors contributed equally to this work}
\affiliation{%
  \institution{Applied Research, Thoucentric}
  \city{Bangalore} 
  \country{India}
}
\email{manujoseph@thoucentric.com}

\author{Harsh Raj}
% \authornote{These authors contributed equally to this work}
\authornotemark[1]
\affiliation{%
  \institution{Applied Research, Thoucentric}
  \city{Bangalore} 
  \country{India}
}
\email{harshraj@thoucentric.com}

\author{Anubhav Yadav}
% \authornote{These authors contributed equally to this work}
\authornotemark[1]
\affiliation{%
  \institution{Applied Research, Thoucentric}
  \city{Bangalore} 
  \country{India}
}

\author{Aaryamann Sharma}
% \authornote{Dr.~Trovato insisted his name be first.}
\affiliation{%
  \institution{Applied Research, Thoucentric}
  \city{Bangalore} 
  \country{India}
}
\email{aaryamannsharma@thoucentric.com}

% \renewcommand{\shortauthors}{B. Trovato et al.}

\begin{abstract}
Querying databases for the right information is a time consuming and error-prone task and often requires experienced professionals for the job. Furthermore, the user needs to have some prior knowledge about the database. There have been various efforts to develop an intelligence which can help business users to query databases directly. However, there has been some successes, but very little in terms of testing and deploying those for real world users. In this paper, we propose a semantic parsing approach to address the challenge of converting complex natural language into SQL and institute a product out of it. For this purpose, we modified state-of-the-art models, by various pre and post processing steps which make the significant part when a model is deployed in production. To make the product serviceable to businesses we added an automatic visualization framework over the queried results.
\end{abstract}



\maketitle

\section{Introduction}
With the advancement in data storage and analysis technologies there has been an exponential increase in the amount of data generated on the web. In 2020 it was estimated that 2.5 quintillion bytes of data \cite{quintillion} was generated everyday with 4.66 billion active Internet users worldwide \cite{social_media}. This calls for efficient tools for parsing through the data for a sound analysis. To improve accessibility, most of this data generated is stored in databases of large companies. Various Relational Databases are developed for this purpose, with SQL being widely used. But writing efficient SQL queries is a task to learn and might be a hurdle for a person belonging to non-technical domains. 

Search analytics and AI-enabled solutions  A self-service AI enabled analytics can help business stakeholders to explore data, ask questions and obtain answers to them in the same way as they would run a search on google. They require users to have zero technical training and provide instant access to information with the help of a search bar. These intelligent solutions can access numerous datasets stored in databases, analyze huge volumes of data, spot hidden trends and return personalized insights in a matter of few seconds. 

The first goal of the suggested effort is to close the communication gap between technical and nontechnical users or users not very well versed in writing efficient codes to query large databases using SQL. In other words, facilitating computer-human connection without any programming experience. Similarly, the second goal is to understand and visualize the dynamics of the data being stored in large databases. This is a tough challenge due to the vastness of the data and the complexity of the queries. It is just not possible for nontechnical users to comprehend the visualizations of the data in large databases using SQL. However, using the proposed solution described below, it is possible to overcome these problems. 

The solution proposed is to use a natural language processing (NLP) based method to automatically generate SQL queries from the questions asked in plain English. The results obtained after parsing the generated SQL query are then sent through an automatic visualization pipeline. This advances the product towards a sophisticated tool for AI driven large database analytics.

The major contributions of this paper are:
\begin{enumerate}  
  \item An end-to-end system which goes from natural language queries to retrieved data and automatically generated visualization
  \item A pre-processing system to maximize the effectiveness of the trained model.
  \item A robust post-processing system which provides guard-rails to the model-generated SQL queries and make it useful
  \item An interactive visualization system which starts with automatically generated visualizations and then allow the user to change them as necessary.
\end{enumerate}

\section{Related Work}

\subsubsection{Natural Language to SQL Translation} 
There has been previous work in text-to-SQL systems ranging from rule-based systems to more advanced and accurate techniques involving state-of-the-art algorithms. Prior works like TableQA \cite{https://doi.org/10.48550/arxiv.2006.06434} uses a combination of deep learning-enabled entity extractor and aggregate clause classifier to build the SQL query. The entity extractor matches the conditions by mapping the columns with their respective values if there are any in the question. It uses a pre-trained question answering model trained on SQuAD \cite{rajpurkar-etal-2016-squad} to help easily locate the column's value from the input question. This has an advantage in the usage of heuristics, since the heuristics can be modified in order to improve performance on the solution in most cases without having to re-train the deep learning model. But as the natural language gets more complex, more errors occur in the retrieval of information from the table. 

With the advent of Transformer \cite{https://doi.org/10.48550/arxiv.1706.03762} architectures there has been a rapid growth in sequence to sequence \cite{article} modelling which is the motivation for text-to-SQL systems. To build such systems it needs to have a joint reasoning of the Natural Language utterances and the structured database schema information. These systems can be recognized as a type of domain-specific semantic parsers. 

There is a class of methods where querying tabular data is done without generating logical forms and targets to get the answer to the question directly from the model. One such method is TAPAS \cite{Herzig_2020}, which extends the Masked Language Modeling (MLM) approach to structured data. Like BERT \cite{https://doi.org/10.48550/arxiv.1810.04805}, TAPAS \cite{Herzig_2020} uses features which have to encode the tabular input. It then initializes a joint pre-training of text sequences and tables trained end to end and is successfully able to restore masked words and table cells. It can infer from a query by selecting a subset of table cells. If available, aggregate operations are also performed on top of them. However, it comes with concerns over memory issues and expensive compute requirements because the entire table has to be fed in as context.

The other paradigm uses an intermediate logical representation (like SQL) to go from natural language queries to its corresponding logical forms and then use the logical form to query the answer. Recent work \cite{https://doi.org/10.48550/arxiv.1905.08205, https://doi.org/10.48550/arxiv.1911.04942, https://doi.org/10.48550/arxiv.1905.06241} shows that using powerful pre-trained language models can further improve these highly specific parsers, even though these language models are trained for pure text encoding. 

There are several datasets for evaluation systems like WIKISQL \cite{https://doi.org/10.48550/arxiv.1709.00103} which has a significantly large number of SQL queries and tables. But all SQL queries are simple, and each database is only a simple table without any foreign key. More such datasets include ATIS \cite{data-atis-original}, Geo \cite{data-geography-original}, Academic \cite{data-academic} but each of them contains only a single database with a limited number of SQL queries, and has exact same SQL queries in train and test splits. Recently released SPIDER dataset spans the largest range in types of queries. It is comparatively complex and cross-domain offering a critical model evaluation towards generalizability. 

The recent state-of-the-art models evaluated on Spider use various attentional architectures for question/schema encoding and AST-based structural architectures for query decoding. IRNet \cite{https://doi.org/10.48550/arxiv.1905.08205} encodes the question and schema separately with Long Short Term Memory (LSTM) \cite{10.1162/neco.1997.9.8.1735} and self-attention respectively. Abstract Syntax Tree (AST) based decoder as described in \cite{yin-neubig-2017-syntactic} is further used to decode a query in an intermediate representation (IR) that exhibits higher-level abstractions than SQL. Previous works has used several encoding schemes including GNN in \cite{https://doi.org/10.48550/arxiv.1905.06241}. These works emphasize schema encoding and schema linking, but design separate featurization techniques to augment word vectors (as opposed to relations between words and columns) to resolve it. Moreover, RAT-SQL \cite{https://doi.org/10.48550/arxiv.1911.04942} framework jointly encodes pre-existing relational structure in the input as well as induced “soft” relations between sequence elements in the same embedding.  

The current state-of-the-art on SPIDER \cite{https://doi.org/10.48550/arxiv.1809.08887} dataset are PICARD \cite{https://doi.org/10.48550/arxiv.2109.05093} and SADGA \cite{https://doi.org/10.48550/arxiv.2111.00653}. SADGA is built on pretrained GAP \cite{https://doi.org/10.48550/arxiv.2012.10309} model which is in turn a modification of RAT-SQL \cite{https://doi.org/10.48550/arxiv.1911.04942} framework. While PICARD \cite{https://doi.org/10.48550/arxiv.2109.05093} is a text-to-SQL semantic parser built upon pre-trained encoder-decoder models. It constrains the decoder through SQL Abstract Syntax Tree (AST) to produce valid SQL queries after fine-tuning the model on the text-to-SQL task.   

\subsubsection{Automatic Visualization} 
Automatic visualization is a widely researched topic in analytics, but introducing robust and efficient automation has always been a hurdle for researchers.  This is hard, if not impossible, since among numerous issues, no consensus has emerged to quantify the goodness of a visualization that captures human perception. 

% Overall, advances in Machine Learning have to some extent helped solve this. We use a tool which adapts binary classifiers for visualization recognition to determine the meaningfulness of a specific visualization. Moreover, it trains a specific ranking model to rank visualizations produced from different combinations of dataset columns and visualization types.

Efforts to produce automatic visualization tools started with building semi-automatic tools with minimal human-interactions such as SAGE \cite{10.1145/223355.223751} and BDVR \cite{10.1145/1502650.1502695}. Text-to-Viz \cite{https://doi.org/10.48550/arxiv.1907.09091} and Click2Annotate \cite{5652885} are designed for no-human interaction and efficiently recommend visualizations. Tools built on rule-based algorithms are prone to errors, but with the success of machine learning reliable systems are possible. With proper combination of machine learning, visual elements and user defined rules \cite{Chen_2019} developed automated infographics.  

Turning the visuals into vector representations is the foremost challenge in efficient usage of machine learning techniques. \cite{8805442} listed some challenges of using machine learning in automatic visualization setup. It concluded that it is difficult and error-prone to organize visual’s characteristics into unique vectors. However, Deepeye \cite{8509240} led improvements to standardize featurization process. It produces representations which are simple yet effective to drive machine learning algorithms. It uses these features to learn to rank from a pool various probable visualization a dataset can produce. Recent techniques also try to model automatic visualization as a recommendation system \cite{10.1145/3313831.3376880}. The problem objective of ranking task is simpler than building a recommendation system whose data collection is difficult. Additionally, recommendation systems perform best with online learning which is difficult to deploy. 

\section{Solution Overview}
\begin{figure*}[!ht]
    \centering
    \includegraphics[width=0.9\linewidth]{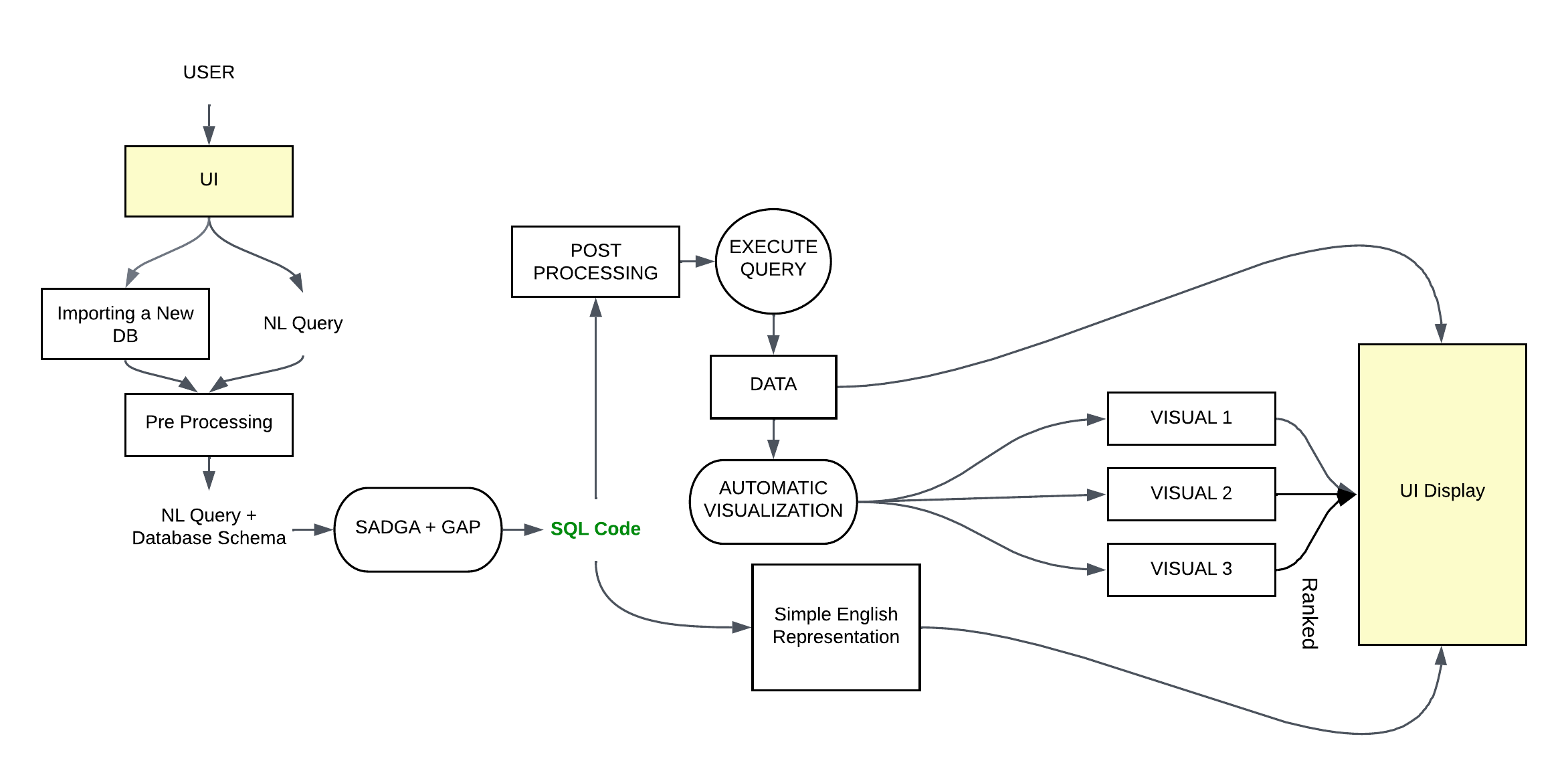}
    \caption{End-to-end System}
    \label{fig:endtoend}
\end{figure*}

We have designed an end-to-end system (Figure: \ref{fig:endtoend}) using state-of-the-art methods in processing natural language to SQL and automated visualization to develop a user-ready product. The key differentiating factor of our approach is the ability to translate business queries into data and visualizations automatically. 

State-of-the-art systems for translating natural language to SQL are still not usable out-of-the-box. They still need quite a bit of hand holding by ways of pre-processing the inputs and post-processing the outputs to have better user experience and reliability. The business usability of the product is also enhanced with the \textit{Automatic Visualization} tool built to analyze the results extracted from the database. The product is an integrated solution that can be easily deployed by businesses without the need for additional training or support. The search and query algorithms are optimized for easy scalability to businesses working with large databases. The product is deployed with a simple yet sophisticated interface for the users and it is an easily accessible solution that can be used by businesses, marketing firms, and data scientists to perform rapid market study, analysis, and real time querying the database with just a few lines of Language.

The product is composed of the following three modules: 

1. Natural Language to SQL Translation 

2. Automatic Visualization  

3. User Interface Module  

The first module is responsible for connecting the user uploaded database and the Natural Language query to the Machine Learning pipeline. The deployed model then produces the resulting SQL code and the answer with a simple parsing of the database. The next module deals with analyzing the result using autonomous visualization techniques. The visualization method stands out because of its ability to automatically detect the type of data and relationship between different attributes. This module also detects the outliers and produces visualizations accordingly with an intelligence built to rank various possible types of visuals (like graphs and charts) based on insightfulness. The User Interface Module is the last module which takes care of providing an interface to the user which is both simple and sophisticated at the same time. It is designed keeping in mind the user experience and business needs. The user interface is interactive and allows the user to perform various tasks such as uploading database, querying the database, and visualizing the results. 

\subsection{Natural Language to SQL Translation}

In our experiments we used variants of SADGA \cite{https://doi.org/10.48550/arxiv.2111.00653} model and experimented it variants – SADGA + Glove, SADGA + Bert-large and SADGA + GAP. As reported in their paper and observed from our experiments we concluded that SADGA + GAP has the highest accuracy in producing correct SQL queries. SADGA \cite{https://doi.org/10.48550/arxiv.2111.00653} adopt the graph structure to provide a unified encoding model for both the natural language question and database schema. Based on the proposed unified modelling, SADGA \cite{https://doi.org/10.48550/arxiv.2111.00653} devise a structure-aware aggregation method to learn the mapping between the question-graph and schema-graph. The structure-aware aggregation method is featured with Global Graph Linking, Local Graph Linking, and Dual-Graph Aggregation Mechanism. To initialize the input embedding of question words and tables/column, the authors use pre-trained embedding techniques. Glove \cite{pennington-etal-2014-glove} is a common choice for embedding initialization whereas BERT, a transformer-based framework is a mainstream embedding initialization method. Moreover, domain specific pre-trained models like GAP \cite{https://doi.org/10.48550/arxiv.2012.10309} is also applied to take advantage of the prior text-to-SQL knowledge. GAP \cite{https://doi.org/10.48550/arxiv.2012.10309} is a model pre-training framework that jointly learns representations of natural language utterances and table schemas. It is trained on 2M utterance-schema pairs and 30K utterance-schema-SQL triples, whose utterances are produced by generative models. SADGA +GAP achieved 2nd place on the challenging Text-to-SQL benchmark SPIDER \cite{https://doi.org/10.48550/arxiv.1809.08887} at the time of writing. However, variants of PICARD \cite{https://doi.org/10.48550/arxiv.2109.05093} model is on the top of the SPIDER \cite{https://doi.org/10.48550/arxiv.1809.08887} benchmark we didn’t use it for our case because in comparison to models like SADGA \cite{https://doi.org/10.48550/arxiv.2111.00653} and GAP \cite{https://doi.org/10.48550/arxiv.2012.10309} it is difficult to setup and is heavier to run inference, also it does not add much accuracy improvements w.r.t. the requirements it adds up. 

After comparative study and evaluation, we found that our chosen model – SADGA + GAP \cite{https://doi.org/10.48550/arxiv.2012.10309} is competitive on industry standards queries and databases. We probed the model on various complex queries and databases which use a mix of various operators like JOIN, ORDER BY, GROUP BY, etc. Overall, the models performed well on most of the queries but failing on more advanced types of queries like the one using Window functions. We chose SADGA \cite{https://doi.org/10.48550/arxiv.2012.10309} with a GAP \cite{https://doi.org/10.48550/arxiv.2012.10309} backbone because not only was the accuracy higher, but also it generated more succinct SQL queries.

% \subsection{Non-Linear Decision Trees}
% \begin{figure*}[!h]
% \centering
% \begin{subfigure}{.4\textwidth}
%   \centering
%   \includegraphics[width=1.2\linewidth]{images/Pre-Processing.jpeg}
%   \caption{Pre-Processing Work Flow}
%   \label{fig:Pre-Processing}
% \end{subfigure}%
% \begin{subfigure}{.6\textwidth}
%   \centering
%   \includegraphics[width=1.1\linewidth]{images/Post-processing-v2.jpeg}
%   \caption{Post-Processing Work Flow}
%   \label{fig:Post-Processing}
% \end{subfigure}
% \caption{Differentiable non-linear decision trees for a dataset with 3 features (Single Layer)}
% \label{fig:dndt_single}
% \end{figure*}

\subsection{Pre Processing}
\begin{figure*}[!ht]
    \centering
    \includegraphics[width=0.9\linewidth]{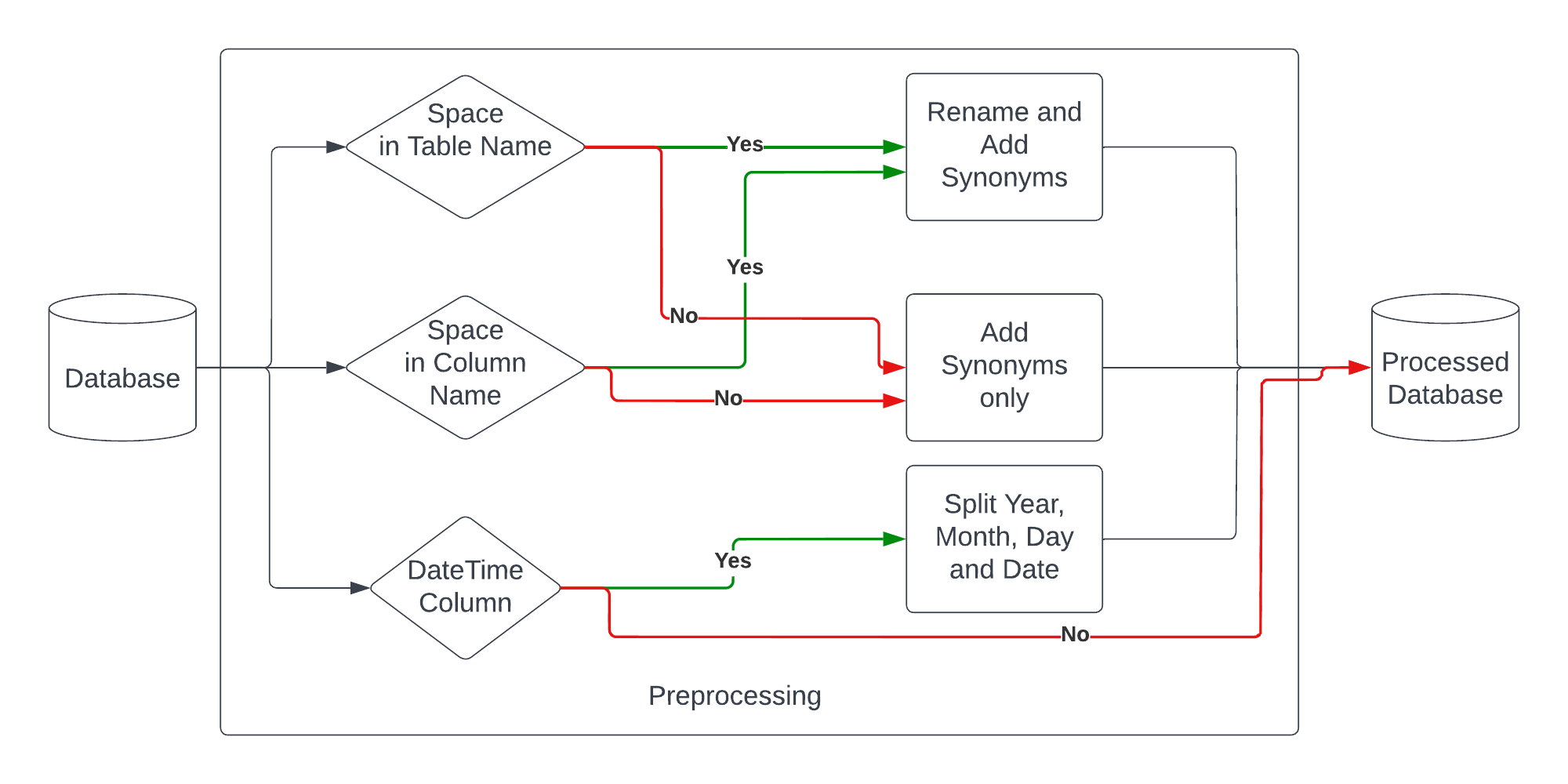}
    \caption{Pre-Processing Work Flow}
    \label{fig:Pre-Processing}
\end{figure*}

The underlying LLM-based model which translates English queries to SQL queries, while being state-of-the-art, still has some rough edges which makes it difficult to use it in an application without some guardrails. We have designed a pre-processing pipeline (Figure \ref{fig:Pre-Processing}) that smooths out the rough edges. These pre-processing steps is strictly related to the initial on-boarding of a dataset and is a one-time activity. 

\subsubsection{Enforce English Schema}
We rely on large language models (LLM) as a backbone to derive semantic relationships between the query and the schema and construct a SQL query. And the LLM’s are typically trained in English. Therefore, it is important for the SQL schema to be as close to English as possible. Our experiments also confirmed the fact that SQL schema being close to English and having meaningful column names drastically impact the accuracy of the conversion. We ensure that any dataset that is on-boarded into the system has plain English column names and those column names rightly represent the content of those columns. For instance, a SQL table of invoices has a column called, \emph{PCs} which denote the number of items ordered. Our pre-processing would change it to \emph{quantity} so that the column name rightly represents the content of the column. This is conceived as a manual process which needs to be done once on on-boarding a new database into the system.

\subsubsection{Boost Semantic Meaning using Synonyms}
Although the LLM based approach try to capture semantically similar words, we found that concatenating meaningful synonyms to a column name increases the accuracy of the underlying model.  For instance, if we have a column named heading, we can add title, and headline to the column name so that the new column name is \emph{heading\_title\_headline}. But before introducing such synonyms we should also make sure the newly introduced terms do not create ambiguity with other columns in the database. This is another manual process where the data scientists can inject specific domain knowledge into the system.

\subsubsection{Removing Punctuation and Spaces Cleaning the Schema entries}
When there is spaces between column name or table name our model consider it as part of string and will produce SQL query in which table/column name will be as it is (with spaces). And without sufficient safeguards, this will cause syntax errors in SQL execution. For instance, if the column name is \emph{toss winner}, the resulting query would have \texttt{SELECT match.toss winner from match} which would throw a syntax error. Therefore, to avoid such problems, we replace the spaces with underscores and remove all the punctuation to get \texttt{SELECT match.toss\_winner from match}

\subsubsection{Special Processing of Datetime Columns} \label{special_date_processing}
Many databases having datetime information as a timestamp or date. But when the user queries something like \emph{what is the total revenue in year 2020}, the model will struggle to extract the meaning out of randomly formatted datetime column and add the right filter to select the year 2020. During the database on-boarding, our system asks the user about datetime columns and the format in which the datetime is represented. And using this information, the system generates different columns for the day, month, year and so on. If the date column is named \emph{invoice\_date}, these additional columns would be named \emph{invoice\_date\_day}, \emph{invoice\_date\_month}, \emph{invoice\_date\_year}, and so on. To have more variety in the way we capture datetime information, we include separate columns for month in short and long forms so that both Oct and October are recognized equally well. In our experiments, this kind of splitting of the datetime information led to much better responses regarding dates. 

\subsubsection{Runtime Processing of queries for standardizing Datetime}
In spite of the additional datetime columns we created in \ref{special_date_processing}, it was observed that when datetime or variations of date is present in in the natural language query as is, the SADGA model can get confused and gives out erroneous SQL queries. This can be because the model doesn’t have an inherent understanding of time and the various ways humans talk about date and time. For instance, humans can refer to time as an exact date (4th July), or at a higher granularity (June, 2021), or in relative terms (last month). Therefore, we have designed a preprocessing step to capture all such variations and standardize them into a form the model understands. We use \textit{SUTime} from Stanford CoreNLP \cite{manning-etal-2014-stanford} to perform named entity recognition on dates and subsequent conversion of dates into standard python format. Once we have identified the dates, we replace them using one of the two formats: 

If the datetime utterance in the query does not mention the specific date then we replace it with \emph{Month:<Month>, Year: <Year>} 

If the datetime utterance mentions a specific date, then we replace it with the date in \emph{yyyymmdd} format. 

\subsection{Post Processing}
The SQL query that the model generates still needs to be processed to make it usable and explainable to the end user.

\subsubsection{Terminal Word Disambiguation}
\begin{figure*}[!ht]
    \centering
    \includegraphics[width=0.9\linewidth]{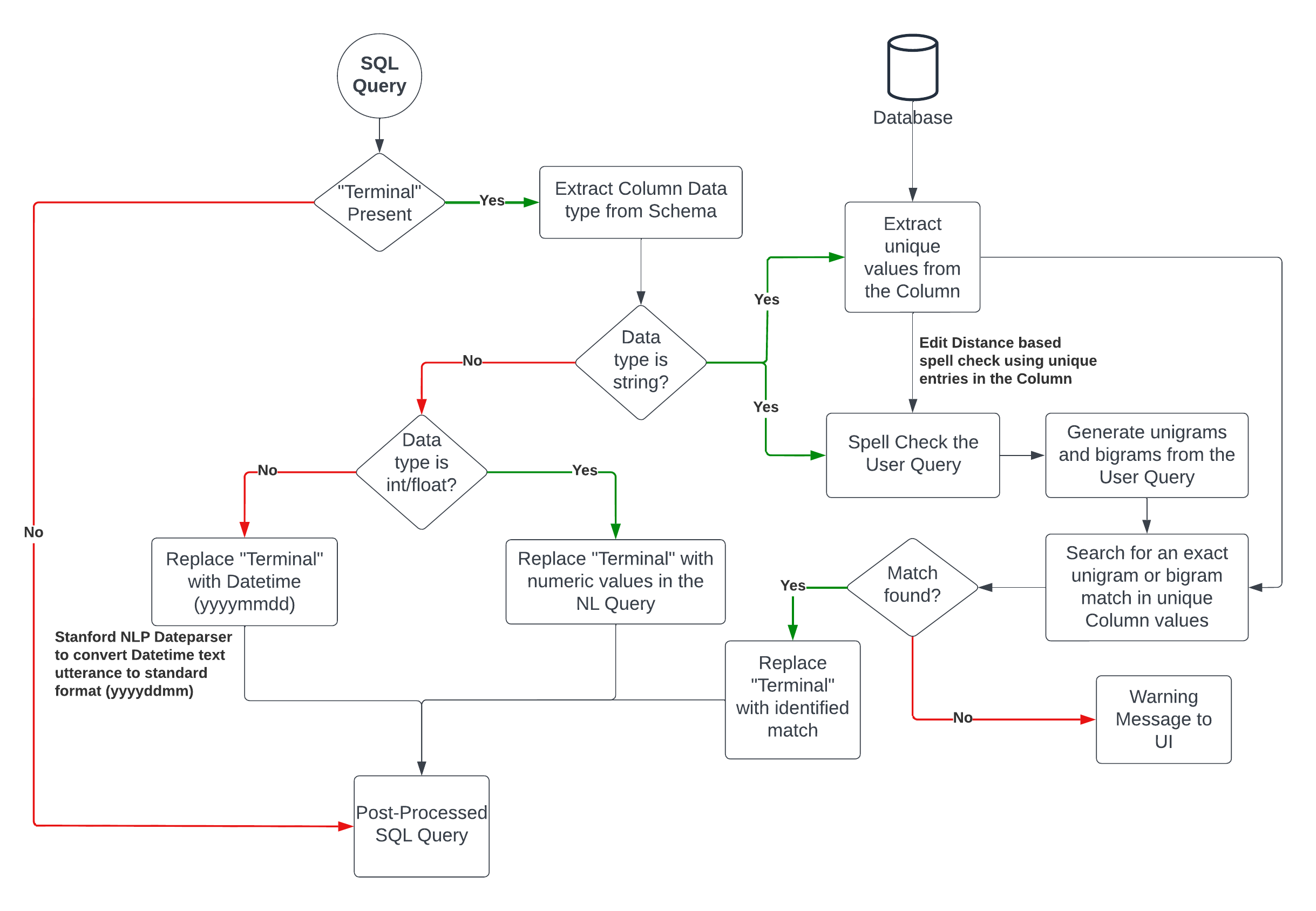}
    \caption{Post-Processing Work Flow}
    \label{fig:Post-Processing}
\end{figure*}

To copy instances of text (like cell values) directly into the generated SQL query the model needs to possess a copying mechanism. Despite a few like finetuned T5 \cite{https://doi.org/10.48550/arxiv.1910.10683}, most of the current state-of-the-art models are not trained to copy cell values from the text. By design the authors of the chosen model have evaluated the model on Component Matching \cite{https://doi.org/10.48550/arxiv.1809.08887} over Exact Match \cite{https://doi.org/10.48550/arxiv.1809.08887}, thus the query contains ambiguous word \emph{Terminal} as a placeholder for the actual cell value. For instance, the natural language query \emph{How many times have earthquakes occur in Colorado?} gets correctly translated into a valid SQL, but instead of Colorado, the generated query has \emph{Terminal} Therefore, we have designed a pipeline (Figure \ref{fig:Post-Processing}) which uses the schema information along with the natural query to replace \emph{Terminal} words with the correct value. 

From the generated SQL query, we first identify all the Terminal’s that we need to replace. And for each terminal, we find out the corresponding column it refers to. This is a trivial task using a SQL Parser \cite{sqlparser} \footnote{\label{sql_parser_link}https://pypi.org/project/sqlparse/}. Once we identify the column and the table, we process the \emph{Terminal} string separately for each data type. 

If the datatype is textual, we use a heuristic-based search to pick the right word from the query to insert in place of \emph{Terminal}. This is not trivial because there are many possible causes of ambiguity. The user may not have spelled the term correctly. The term that we are looking for can be a combination of words instead of a single word.
\begin{enumerate}
    \item We start off by extracting the unique values from the column from the database.
    \item Now both the user query and the extracted unique values go through a bit of basic cleaning where we convert everything to lowercase and remove punctuation.
    \item After the cleaning, we do a spell check based on edit distances between the words in the user query and the words in the English dictionary and words in the unique values extracted from that particular column.
    \item After the spell check, we generate unigrams and bigrams from the user query and do a word matching search with the unique values in the extracted column.
    \item If we find a match, we replace the \emph{Terminal} with the matched value and if not, we send a warning to the UI to urge the user to check the query again. 
\end{enumerate}
% We start off by extracting the unique values from the column from the database. Now both the user query and the extracted unique values go through a bit of basic cleaning where we convert everything to lowercase, remove punctuation and so on. After the cleaning, we do a spell check based on edit distances between the words in the user query and the words in the English dictionary and words in the unique values extracted from that particular column. After the spell check, we generate unigrams and bigrams from the user query and do a word matching search with the unique values in the extracted column. If we find a match, we replace the \emph{Terminal} with the matched value and if not, we send a warning to the UI to urge the user to check the query again. 

If the datatype of the column is numeric, we extract all the numbers from the user query. We also pre-process the user query to convert any numbers which are in the English form to numbers. For instance, if the user query is \emph{show me the number of customers who clicked on the ad at least two times} we pre-process it to make it \emph{show me the number of customers who clicked on the ad at least 2 times}. In cases where there are multiple numbers in the user query and the multiple numeric \emph{Terminal}, the assignment becomes non-trivial. After experimenting with multiple ways of resolving the ambiguity, we found that the order in which the numbers appear in the user query will be the same order in which the SQL query would be generated by the model. Therefore, keeping that principle in mind, we assign the numeric values using the order that it occurred in the user query. 

If the datatype of the column is \emph{datetime}, we replace the terminal value with date in a standard datetime format \emph{yyyymmdd}. The datetime text utterance is converted to the standard format through pre-processing (sec) 

\subsubsection{Simplified and User-Friendly Representation for SQL}
In addition to the data we retrieve using the SQL query, we also wanted to show the salient points from the SQL query in plain English without the syntax of SQL. Instead of relying on heuristics based on string matching the raw SQL query, we first used a SQL parser \cite{sqlparser} which does syntax-aware parsing of the SQL query into different tokens like \textit{DML}, \textit{Whitespace}, \textit{Identifier}, \textit{Keywords}, \textit{Where}, etc. And over these parsed tokens, we implemented a heuristic to reduce the SQL query into something more simpler: \begin{enumerate}
    \item Using the schema, we pick up all the table names which the query is addressing 
    \item Conditions in the SQL query are extracted using the Where keyword 
    \item If keywords like \texttt{INTERSECT} and \texttt{UNION} are present in the, we split the query in two and use the same process to generate English representation for each part and then join using intersect or union. 
\end{enumerate}

The whole system works to extract key information from the SQL query and represent it in a format that is easy for a non-technical user to read and understand. For instance the query - \texttt{SELECT * from customers where customers.region = 'INDIA' – would be represented as Column(s): All Table(s): customers, Filtered on: customers.region = 'INDIA'}. 

\subsubsection{Automatic Visualization}
We enable the user with an automatic visualization tool on top of the data the SQL query fetches from the database. For this we have used an open-source software, Deepeye \cite{8509240} which organizes different visualization charts and graphs into graph-based structure with a specific hierarchy based on their rank. The query result is passed to the automatic visualization pipeline based on Deepeye  \cite{8509240}. The performance of machine learning algorithms depends heavily on the choice of features. Using metrics like correlation, ratio, etc, features are built on a given combination of columns and visualization types. These features are represented in a graph with each node representing a visualization. Now Deepeye \cite{8509240} ranks them based on their quality. Deepeye \cite{8509240} uses a ranking based model which is trained with a set of labeled data. Deepeye \cite{8509240} supports three types of ranking approaches -  
\begin{itemize}
    \item \textbf{Learning-to-rank}: It is a supervised learning task that takes the input space X as lists of feature vectors, and Y the output space consisting of grades (or ranks). The goal is to learn a function $F(·)$ from the training examples, such that given two input vectors $x1$ and $x2$, it can determine which one is better, $F(x1)$ or $F(x2)$. 
    
    \item \textbf{Partial Order-based approach}: it defines some partial orders which are used to decide which visualization node is better. A graph based on the partial orders is built, where each vertex is a visualization node and the directed edge between two nodes are decided by the partial order. Finally, the graph is used to compute a score for each visualization node based on topology sorting, i.e., the smaller the topology order is, the larger the score is.  
    
    \item \textbf{Diversified Ranking}: it selects diversified top-k visualization nodes since there may be many similar visualizations showing redundant information. For example, $v1 > v2$ and $v2 > v3$ do not necessarily mean that $v1 + v2 > v1 + v3$, since $v1$ and $v2$ might be very “similar". It mainly constructs a graph in which nodes are visualizations, and weight of the edge between two nodes denotes the distance between them. Then the graph is used to define relevance and diversity measurement to calculate diversified top-k visualizations. 
\end{itemize}

After some experiments with all the above type of techniques we found that \emph{Diversified Ranking} is the most apt for this task as it is better in producing visuals of varied range, hence useful for multi-purpose analytics. 

One of the major challenges in automatic visualization is the ambiguity in the user intent. There are many ways a particular data can be plotted and to exactly specify what is in the user’s mind is typically challenging. Although DeepEye \cite{8509240} goes a long way towards making those decisions automated, we still may have ambiguity. What each user wants is subjective to his views and aesthetics. For instance, the query returned data of total revenue per region for the last year. And suppose, DeepEye \cite{8509240} suggested a pie chart as the best visualization for this data. But what if the user hates pie-charts and wants to view it as a bar chart? On the other hand, if we ask the end-user to decide what visualization he wants to generate, it will put unnecessary cognitive load on him as well. Therefore, we have taken a hybrid approach, where we use DeepEye \cite{8509240} to generate the most likely visualization and then leave the control to the user where he can either cycle between the different options suggested by DeepEye \cite{8509240} or even take matters into his own hands and select the kind of visualization he wants. 

\subsubsection{User Interface Module}
We have developed the user interface as a web application using Django framework \cite{django}. There are a few databases pre-loaded, but the user interface (UI) also enables the user to upload various data sources in SQLite, \emph{csv} or \emph{xlsx} file formats (multi-table databases are only supported in SQLite) through the \textit{Upload page}. The upload process will also carry out the sufficient pre-processing and tokenization of the schema to make it ready for inference. Once the upload finishes, the user will get a notification and the new database is ready to use. 

\begin{figure*}[!h]
    \centering
    \includegraphics[width=0.8\linewidth]{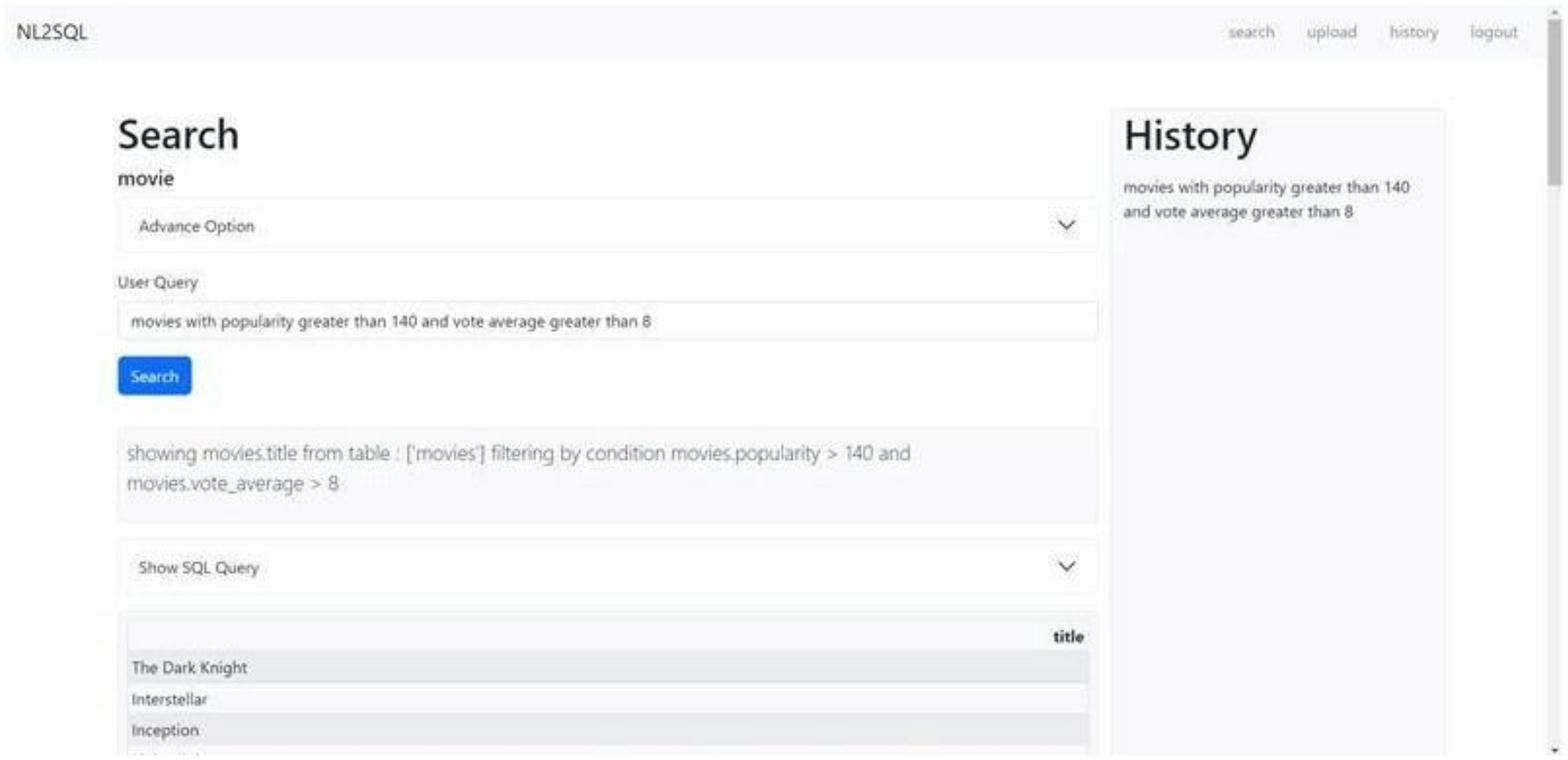}
    \caption{User Interface}
    \label{fig:ui}
\end{figure*}

The user can then head to the \emph{Search} page and select the dataset he wants to analyse. Using a simple search box, we enable the user to ask questions in English which is then relayed to the backend system which generates the right SQL query for the English query by the user. This query is then used to retrieve the data from the database. This data is displayed as a table in the UI and along with it the automatically generated visualization is also displayed to the user. The user can also choose between different visualization options based on their preference. The user can then download the results generated by the SQL query in form a downloadable \emph{csv} by clicking on the download button next to the results. The user can also save the visualizations generated by right clicking on charts and saving them to their system.  We are also saving all the user searches and the corresponding queries in a cache so that we need not do model inferencing for queries which the model has already generated the query for. The history of user searches is also found in the \emph{History} page as well as in the side bar on the \emph{Search} page. The visual of the UI can be seen in Figure \ref{fig:ui}.

\subsection{Evaluation}
Since a quantitative evaluation for the end-to-end system wasn't possible, we present key insights of both success and failure models in Table \ref{eval_table}. Most of the normal use-cases the system is able to handle and generate the right queries.Misspelled queries (Jhonny instead of Johnny) are handled by the system in a robust manner (Query 4 in Table \ref{eval_table})

\begin{table*}[!h]
\centering
\caption{Sample Queries processed through the system}
\begin{tabular}{p{6cm}p{6cm}p{5cm}} 
\toprule
\textbf{QUERY}                                                                                                                      & \textbf{SQL CODE}                                                                                                                                                                                                              & \textbf{COMMENTS}                                                                                                          \\ 
\hline
Which brand car has the most customers?                                                                                             & SELECT Brands.brand\_name FROM Dealer\_Brand JOIN Brands ON
  Dealer\_Brand.brand\_id = Brands.brand\_id GROUP BY Brands.brand\_name ORDER BY
  Count(*) Desc LIMIT 1                                                          & The query is right                                                                                                         \\ 
\hline
What are the email addresses of the
  customer whose first name is MARY?                                                            & SELECT customer.email FROM customer WHERE customer.first\_name = 'MARY'                                                                                                                                                        & The query is right                                                                                                         \\ 
\hline
Name all movies starring Johnny
  Cage OR Name all movies starring Jhonny Cage                                                      & SELECT film.title FROM actor JOIN film JOIN film\_actor ON actor.actor\_id
  = film\_actor.actor\_id AND film\_actor.film\_id = film.film\_id AND
  actor.actor\_id = film\_actor.actor\_id WHERE actor.first\_name = 'JOHNNY' & Was able to identify Johnny even though misspelled, but wasn't able to
  link "Cage" to the last name in the schema  \\ 
\hline
Give the top 2 places with maximum depth
  of earthquake                                                                            & SELECT quakes.place FROM quakes ORDER BY quakes.depth Desc LIMIT 1                                                                                                                                                             & Was able to generate the right query, but selected top 1 instead of 2                                                      \\ 
\hline
\begin{tabular}[c]{@{}l@{}}Which places had a positive longitude value? OR~\\ Which places in the western hemisphere?~\end{tabular} & SELECT DISTINCT quakes.place FROM quakes WHERE quakes.longitude =
  'Terminal'                                                                                                                                                 & The system does not have an understanding of complex concepts like
  positive (0) or western and eastern hemispheres       \\ 
\hline
Which team won the most number of
  matches?                                                                                        & SELECT matches.team1 FROM matches GROUP BY matches.team1 ORDER BY
  Count(*) Desc LIMIT 1                                                                                                                                      & The query is right                                                                                                         \\
\bottomrule
\end{tabular}
\label{eval_table}
\end{table*}

\subsection{Discussion and future work}
\subsubsection{Current Problems} 
For a reliable deployment of the SOTA Machine Learning techniques for SQL to Natural Language translation it needs to be robust to wide range of advanced SQL structures which currently the models are uncertain. We introduced various pre- and post-processing steps to the system and added a human-computer interface to not totally rely on the system and be free from unprecedented failure. We also considered the business perspective of such systems and added Automatic visualization techniques over it with a User Interactive Interface. Below are few areas where the current systems lack -  
\begin{enumerate}  
    \item The system has some difficulty in completely picking up the required filtering as evidenced by Query 4 in Table \ref{eval_table}
    
    \item The model does have an affinity towards more popular use cases like \textit{Top 1}, even though the query mentioned \textit{Top 2} (Query 5 in Table \ref{eval_table}. But these are sporadic instances.
    
    \item For larger databases (with many tables and relationships) the system is not effectively able to produce results, this might be because of question the user is giving is difficult to understand as many synonyms can be used to for the name of column or feature of database 
    
    \item The system has no understanding of complex concepts as evidenced by Query 6 in Table \ref{eval_table}.
    
    \item It is not able to process advanced SQL queries such as those which contains \emph{Window functions} like \texttt{PARTITION BY}, \texttt{RANK} and \texttt{MERGE}. 
    
    \item The system is not constrained to produce efficient SQL codes for memory and time complexity optimization. 
    
    \item Time complexity of the terminal disambiguation is a challenge. 

\end{enumerate}  

\subsubsection{Possible Solutions and future work}
Below are few possible possible solutions to the above issues and future work directions -
\begin{enumerate}
    \item More robust models which can perform a wider variety of SQL operations efficiently. 
    \item Efficient copying mechanism in the query generation phase to generate terminal values in an end-to-end fashion.
    \item Reducing the time complexity of Terminal Disambiguation.
\end{enumerate}
% \newpage
% \pagebreak
% \clearpage
\vfill
\bibliographystyle{ACM-Reference-Format}
\bibliography{sample-bibliography} 

\end{document}